      [1999/12/01 v1.4c Il Nuovo Cimento]
\documentclass[cite]{cimento}
\title{%
The Lifshitz-Khalatnikov Kasner index parametrization and the Weyl Tensor
}
\author{%
Donato Bini\from{ins:C}\from{ins:I}, 
Christian Cherubini\from{ins:B}\from{ins:I},
        \atque
Robert T. Jantzen\from{ins:V}\from{ins:I}
}
\instlist{
\inst{ins:C} Istituto per le Applicazioni del Calcolo ``M. Picone,'' CNR I--00161 Rome, Italy
\inst{ins:I}   International Center for Relativistic Astrophysics -- I.C.R.A.,
  University of Rome ``La Sapienza,'' I--00185 Rome, Italy
\inst{ins:B} University Campus Bio-Medico of Rome, 
   PRABB, Via \'Alvaro del Portillo 21, I--00128 Rome, Italy
\inst{ins:V} Department of Mathematical Sciences, Villanova University, Villanova, PA 19085,  USA
}
\PACSes{\PACSit{00.00}{}
\PACSit{04.20.Cv}{}}

\usepackage{graphicx}  
\usepackage{latexsym}
\usepackage{amssymb}

\input prepictex \input pictex \input postpictex
\def\mathput#1{\relax \ifmmode \displaystyle #1\else $\displaystyle #1$\fi}

\def\fl{}
\def\II{I} \def\JJ{J}  

\def\beq{\begin{equation}}
\def\eeq{\end{equation}}

\def\rmd{{\rm d}}
 
\def\dual{{}^{*}\kern-1pt }

\DeclareSymbolFont{AMSb}{U}{msb}{m}{n}
\DeclareMathSymbol{\R}{\mathbin}{AMSb}{"52}
\DeclareMathSymbol{\I}{\mathbin}{AMSb}{"49}
\DeclareMathSymbol{\C}{\mathbin}{AMSb}{"43}

\def\selfd{\,\mathop{\tilde{\,}}\nolimits\,}
\def\leftselfd{\selfd\kern-1.7pt}

\def\version{\today}

\begin{document}

\maketitle

\begin{abstract}
The scale invariant Petrov classification of the Weyl tensor is linked to the scale invariant combination of the Kasner index constraints, and the Lifshitz-Khalatnikov Kasner index parametrization scheme turns out to be a natural way of adapting to this symmetry, while hiding the permutation symmetry that is instead made manifest by the Misner parametrization scheme. While not so interesting for the Kasner spacetime by itself, it gives a geometrical meaning to the famous Kasner map transitioning between Kasner epochs and Kasner eras, equivalently bouncing between curvature walls, in the BLK-Mixmaster dynamics exhibited by spatially homogeneous cosmologies approaching the initial cosmological singularity and the inhomogeneous generalization of this dynamics. 
\end{abstract}


\section{Introduction}

It is amazing that after years of people using and working with a piece of well known mathematics, there is still the possibility of realizing something new about it that somehow escaped notice for so long. While not earth shaking, the present remarks about the connections between the Weyl tensor and the Kasner index parametrization are certainly interesting. Discovered serendipitously by playing around with 
the speciality index~\cite{BC}, a little thought uncovered some cute mathematical structure \cite{bcj2007}, aspects of which have played a key role in the lore of general relativity in the past half century.

The speciality index $S$ is a  dimensionless curvature invariant which indicates the the special or non-special Petrov algebraic  behavior of the Weyl curvature tensor of a spacetime which is not too degenerate in the Petrov classification of such tensors.
This quantity, originally introduced and used in numerical relativity in the study of  a 
black hole  radiating gravitational waves~\cite{BC,beetle}, also found application to the Petrov classification of perturbed spacetimes~\cite{cbbp1,bertietal}, to wave extraction in numerical relativity~\cite{nerozzietal} and in cosmology~\cite{cbbp2}.  The Kasner spacetime itself has been used as an example for discussing the determination of quasi-Kinnersley frames~\cite{bbb}.

The 1-parameter family of Kasner spacetimes \cite{kasner} is most simply represented in terms of the Kasner indices, three exponents satisfying two identities, and which describe the scalefree part of the eigenvalues of the extrinsic curvature tensor for the spatial geometry. The spatial metric isometry classes correspond to the equivalence classes of Kasner triplets under permutations of the three Kasner axes, leading to a natural action of the symmetric group $S_3$ of permutations of the integers $(1,2,3)$. The Petrov classes characterizing the Weyl curvature tensor involve the scalefree part of the eigenvalues of a complex matrix arising from the self-dual combination of the Weyl tensor and its dual. These classes correspond to equivalence classes modulo the same group action of $S_3$. It is not surprising that there should be a nice relationship between the two mathematical structures, and it is this relationship which gives added meaning to the Lifshitz-Khalatnikov parametrization of the Kasner indices, introduced in order to describe collisions with curvature potential walls in asymptotic gravitational dynamics in a simple way, now seen in terms of simple transitions of the Weyl tensor eigenvalue ratios.

To summarize the key points which the following discussion illuminates in detail:
\begin{enumerate}
  \item 
As described in Ref.~\citen{ES}, the Weyl tensor $C_{\alpha\beta\gamma\delta}$ ($\alpha,\beta=0,1,2,3$) can be repackaged as a complex self-dual tensor $C_{\alpha\beta\gamma\delta}+i \dual C_{\alpha\beta\gamma\delta}$ which in turn with a choice of unit timelike vector field $u^\alpha$ can be repackaged as a tracefree symmetric spatial tensor $Q_{ab}=-(E_{ab}+i B_{ab})$  ($a,b=1,2,3$), which is a complex combination of the electric and magnetic parts of the Weyl tensor with respect to that vector field, and whose eigenvalues $\lambda_a$ therefore sum to zero: $\lambda_1+\lambda_2+\lambda_3=0$, leaving only two independent eigenvalues, and modulo rescaling, only one independent complex number to describe its scalefree part (at a given spacetime point). Any one of the 6 possible ratios $\mu_{ab}=\lambda_a/\lambda_b$ of the three eigenvalues can be taken to parametrize this degree of freedom, leading to the action of the symmetric group $S_3$ on the resulting parametrizations. The classification of the eigenvalues (modulo rescaling) is the Petrov classification.
  \item The family of Bianchi type I Petrov type I (and D) vacuum Kasner spacetimes is parametrized by three Kasner indices $p_a$ which are the eigenvalues of the scalefree part of the extrinsic curvature tensor $K^a{}_b$ satisfying two identities whose scalefree combination is simply $1/p_1+1/p_2+1/p_3=0$, and hence their real ratios $u_{ab}=p_a/p_b$ (six possibilities, with $S_3$ again acting on the choices) parametrize the scalefree part of these eigenvalues, only one of which is independent due to the second independent identity.
  \item For Kasner, the Weyl tensor ratios $\lambda_a/\lambda_b$ agree with the Kasner index ratios $(1/p_a)/(1/p_b)= p_b/p_a$: $\mu_{ab}=u_{ba}$, thus tying the Kasner map at the heart of BKL-Mixmaster dynamics
directly to simple transitions in the scale invariant part of the Weyl tensor \cite{bcj2007}.
\end{enumerate}

After expanding on these points, we review the role of the Lifshitz-Khalatnikov parametrizations $u_{ab}$ of the Kasner indices in the BLK-Mixmaster dynamics \cite{LL,LK0,LK,BK0,BK,BK2,misner69,ryanshepley,unified,bergeretal,lars}.

\section{The Weyl tensor}

Consider a timelike congruence of world lines 
with  unit tangent vector $u^\alpha$ satisfying $u_\alpha u^\alpha=-1$. 
$u^\alpha$ can be interpreted as the $4$-velocity field of an observer family and used to measure tensor fields in the $1+3$ spacetime splitting which follows from the orthogonal decomposition of each tangent space into a local time direction and a local rest space.
This description of spacetime geometry is often referred to as gravitoelectromagnetism~\cite{ellis1,ellis2,mfg,maartens}.

For example, the $(1+3)$-splitting of the self-dual representation
$z_{\alpha\beta}=F_{\alpha\beta}+i{}^*F_{\alpha\beta}$  of a 2-form $F_{\alpha\beta}$
identifies the complex (spatial) vector\footnote{A spatial quantity with respect to $u^\alpha$ is such that any contraction with $u^\alpha$ gives 0.}
\beq
z_\alpha=z_{\alpha\beta}u^\beta
=E_\alpha+iB_\alpha\,, \qquad 
z_\alpha u^\alpha=0\,,
\eeq
where
$E_\alpha$ and $B_\alpha$ denote the electric and magnetic parts of the 2-form with respect to $u^\alpha$, respectively. The relationship between $F_{\alpha\beta}$ and $z_\alpha$ basically establishes an isomorphism between the Lie algebra of the Lorentz group and the Lie algebra $o(3,C)$ of the complex orthogonal group in three dimensions. Completing $u$ to an orthonormal frame $E_\alpha$ ($\alpha=0,1,2,3$) with $E_0^\alpha=u^\alpha$, in terms of which the complex vector $z_\alpha$ only has spatial components $z_a$ ($a,b=1,2,3$), then Lorentz transformations of this frame result in a complex rotation of this complex 3-vector.

Similarly, the self-dual part of the Weyl tensor 
$ (C+i{}^*C)_{\alpha\beta\gamma\delta}$ 
is associated with the complex (spatial) tensor \cite{ES}
\beq
-Q_{\alpha\beta}=  (C+i\,{}^*C)_{\alpha\beta\gamma\delta} u^\gamma u^\delta=
E_{\alpha\beta}+iB_{\alpha\beta}\,,
\eeq
where now the spatial tensors $E_{\alpha\beta}$ and $B_{\alpha\beta}$ denote the electric and magnetic parts of the Weyl tensor associated with $u^\alpha$, respectively.
The spatial tensor $Q_{\alpha\beta}$ is symmetric and trace-free, i.e., satisfies the conditions
\beq
Q^\alpha{}_\alpha=0\,, \quad  
Q_{\alpha\beta}=Q_{\beta\alpha}\,, \quad 
Q_{\alpha\beta}u^\beta=0
\eeq
and its nonzero components in a frame adapted to $u^\alpha$ are a complex symmetric tracefree $3\times 3$ matrix $(Q_{ab})$ $(a,b=1,2,3)$ when expressed in an orthonormal frame adapted to $u^\alpha$. The mixed form of this spatial tensor with component matrix $\mathbf{Q}=(Q^a{}_b)$ can be thought of as a linear transformation on the representation space of complex 3-vectors, and as such its matrix can be classified by its eigenvalues and eigenvectors.

The eigenvalue problem for this $3$-dimensional matrix
\beq 
Q^a{}_b r^b=\lambda \, r^a\,,
\eeq
is equivalent \cite{ES} to the corresponding problem for the Weyl tensor
\beq
\label{eigenval}
\frac12 C^{\alpha\beta}{}_{\gamma\delta} X^{\gamma\delta}=\lambda X^{\alpha\beta}\,,
\eeq
where $X^{\alpha\beta}$ are the eigenbivectors of the Weyl tensor associated with the eigenvalue $\lambda$, represented as complex vectors $r^a$ in the self-dual representation. The algebraic type of the matrix ${\mathbf Q}$ provides an invariant characterization of the gravitational field at a fixed spacetime point. This Petrov classification based on the eigenvalues $\lambda_1,\lambda_2,\lambda_3$ of ${\mathbf Q}$, their multiplicities, and the number of linearly independent eigenvectors leads to certain canonical forms for the matrix ${\mathbf Q}$ called the normal forms which are listed in \cite{ES}. At most two eigenvalues are independent due to the tracefree condition on the matrix
\beq
{\rm Tr}\,{\mathbf Q}=\lambda_1+\lambda_2+\lambda_3=0\,.
\eeq
This classification is insensitive to the scale of the eigenvalues, and so only depends on their scalefree part.

The most general Petrov type I corresponds to $\mathbf Q$ being diagonalizable. A canonical or normal frame $E_\alpha$ for this type is one in which the matrix is diagonal
\beq
\label{typeI}
{\mathbf Q}_I=\pmatrix{
\lambda_1 & 0 & 0 \cr
0& \lambda_2 & 0 \cr
0 & 0 &  \lambda_3\cr
}\,. 
\eeq
Type D occurs as the subcase $\lambda_1=\lambda_2$ and permutations. The remaining canonical forms of ${\mathbf Q}$ can be found in \cite{ES}.

The Petrov classification distinguishes between algebraically general spacetimes (type I) and algebraically special ones (types D, II, N, III and O) depending on the degeneracy of eigenvalues and eigenvectors associated with the above spectral problem. One way of  determining whether a given spacetime is general or special involves evaluating the invariants
\beq\fl\quad
\II =\frac12 {\rm Tr} {\mathbf Q}^2 =\frac12 (\lambda_1^2+\lambda_2^2+\lambda_3^2)\,, \qquad
\JJ =\frac16 {\rm Tr} {\mathbf Q}^3 =\frac16 (\lambda_1^3+\lambda_2^3+\lambda_3^3)\,.
\eeq
For Petrov types I and then D and II in the top two levels of the Petrov hierarchy, both invariants $\II$ and $\JJ$ are nonzero, while both vanish for types O, N and III in the lowest level of the three level pyramid.
Any algebraically special spacetime  satisfies
\beq
\II ^3=27 \JJ ^2
\eeq
so except for most degenerate types O, N and III in the Petrov hierarchy, one can introduce the so called ``speciality index" which is a well defined function of the eigenvalues $\lambda_1,\lambda_2,\lambda_3$ by taking the dimensionless ratio (which is therefore independent of the overall scale of the eigenvalues)
\begin{eqnarray}
\label{SPECT}
\mathcal{S}&=&\frac{27\JJ ^2}{\II ^3}
=6\frac{(\lambda_1^3+\lambda_2^3+\lambda_3^3)^2}{(\lambda_1^2+\lambda_2^2+\lambda_3^2)^3}
\end{eqnarray}
having the value 1 for the algebraically special types D and II and obviously invariant under any permutation of the eigenvalues.

This quantity is related to the similar invariant quantity $M$ introduced earlier by McIntosh and Arianrhodtf \cite{MA1990,AM1991,AM1992} by the relation
\beq
\mathcal{S}=(1+M/27)^{-1}\,.
\eeq
They used this quantity to simplify the discussion the geometry of the set of principal null directions of the Weyl tensor described by Penrose and Rindler \cite{PR}, and applied this insight to the Kasner and Curzon metrics.

The tracefree condition $\lambda_1+\lambda_2+\lambda_3=0$ does not constrain the overall scale of the three eigenvalues, but it does lead to only one independent ratio among them. Choosing $\mu =\lambda_1/\lambda_2$, then it follows from the constraint that $\lambda_3/\lambda_2=-(1+\mu )$ and therefore $\lambda_3/\lambda_1=-(1+\mu ^{-1})$, so the six possible ratios $\mu_{ab}=\lambda_a/\lambda_b$ ($a\neq b$) are given by the following formulas in terms of one of them $\mu=\mu_{12}$
\begin{eqnarray}
(\mu_{12},\mu_{32},\mu_{31})
&=&
\left( \frac{\lambda_1}{\lambda_2},\frac{\lambda_3}{\lambda_2}, \frac{\lambda_3}{\lambda_1}\right)
=(\mu,-(1+\mu),-(1+\mu^{-1}))\,,
\nonumber\\
(\mu_{21},\mu_{23},\mu_{13})
&=&
\left( \frac{\lambda_2}{\lambda_1},\frac{\lambda_2}{\lambda_3}, \frac{\lambda_1}{\lambda_3}\right)
=(\mu^{-1},-(1+\mu)^{-1},-(1+\mu^{-1})^{-1})\,,
\nonumber\\
(\lambda_1,\lambda_2,\lambda_3) &=& \lambda_2 (\mu,1,-(1+\mu))\,.
\end{eqnarray}
Thus any scale invariant (i.e., dimensionless) function of the eigenvalues can only depend on $\mu $, which must be the case for the speciality index.
The invariance of the speciality index under the action of the symmetric group $S_3$ will reflect these six possibilities for a choice of the single independent ratio, and the speciality index will therefore be invariant upon replacing $\mu$ by any of the five remaining expressions $\mu_{ab}$ given in terms of $\mu$.

Using the trace-free property of the matrix ${\mathbf Q}$, one can re-express any eigenvalue in terms of the remaining two, so the speciality index is only a function of two independent eigenvalues, which we choose to be $\lambda_1$ and $\lambda_2$
\begin{eqnarray}
\mathcal{S}
&=& 
\frac{27}{4}\frac{\lambda_1^2\lambda_2^2(\lambda_1+\lambda_2)^2}
{(\lambda_1^2+\lambda_2^2+\lambda_1\lambda_2)^3}\,.
\label{SPECT2}
\end{eqnarray}
Since the speciality index is a dimensionless homogeneous function of these two variables, it only depends on their complex ratio,
leading to a convenient parametrization of this function
\beq
\label{SPECT3}\fl\qquad
\mathcal{S}(u)=\frac{27}{4}\frac{\mu^2(1+\mu)^2}{(1+\mu+\mu^2)^3}
 =\frac{27}{4}\frac{[(\mu+\frac12)^2-\frac14]^2}{[(\mu+\frac12)^2+\frac34]^3}
\,,\quad \mu\in \C\,.
\eeq
$\mathcal{S}(u)$ is a complex analytic function of $\mu$ (clearly symmetric under the reflection $\mu+\frac12\to -(\mu+\frac12)$) whose values along the real axis are confined to the closed interval $[0,1]$,  which vanishes in the limit $\mu \to \infty$, and whose poles are the roots  
$\mu=-\frac12\pm i\frac{\sqrt 3}{2}$ of the denominator $\mu^2+\mu+1$, at which values the condition $|\lambda_1|=|\lambda_2|=|\lambda_3|$ holds. Values of $\mu$ on the real axis correspond to the vanishing  of the magnetic part of the Weyl tensor in a canonical frame, the case of a purely electric Weyl tensor.
The poles are the intersection points of the unit circles $|\mu+1|=1=|\mu|$. Indeed these two circles and the line through their two intersection points (a line which can be thought of as the circles of infinite radius with centers on the real axis at positive and negative infinity) divides up the complex plane into six equivalent regions under the action of the symmetric group. The imaginary part of $\mathcal{S}(u)$ is zero exactly on these two circles and on the line through their intersection points plus along the real axis.

\begin{figure} 
\typeout{*** EPS figure 1}
\begin{center}
\includegraphics[scale=0.23,angle=0]{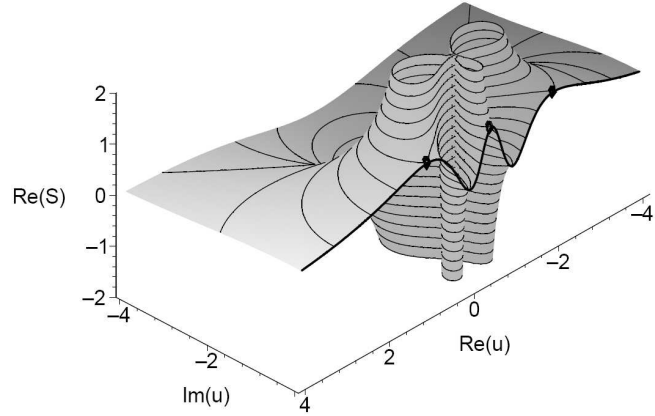} \vglue-2cm 
\end{center}
\caption{%
The real part of the speciality index function $\mathcal{S}(\mu)$ is plotted as a function of the real and imaginary parts of $\mu$. The speciality index values corresponding to vacuum Kasner spacetimes (lying above the real axis) are indicated by a thick solid curve. The Petrov type D cases ($\mathcal{S}=1$) are denoted by bullets on this curve above $\mu=-2,-1/2,1$. } 
\label{fig:1}
\end{figure}

\begin{figure} 
\typeout{*** EPS figure 2}
\begin{center}
\includegraphics[scale=0.23,angle=0]{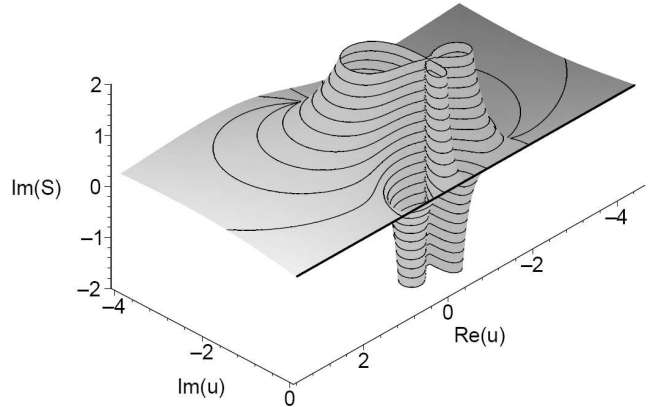}
\end{center}
\caption{The imaginary part of the speciality index function $\mathcal{S}(\mu)$ is plotted as a function of the real and imaginary parts of $\mu$. } 
\label{fig:2}
\end{figure}

In Figs.~\ref{fig:1} and \ref{fig:2} the real and imaginary parts of the speciality index are plotted as functions of the real and imaginary parts of $\mu$. In the real part plot, the speciality index values corresponding to purely electric Weyl tensors lie above the real axis, shown superimposed as a thick curve, and only half the plot is shown since it has even symmetry about the real axis; it also has even symmetry about the line $\Re e(\mu)=-1/2$. In the imaginary part plot, again only half the plot is shown since it has odd symmetry about the real axis; it also has odd symmetry about the line $\Re e(\mu)=-1/2$. Below it will be seen that the real speciality index over the real part of $\mu$ describes exactly the family of Kasner spacetime Weyl tensors.


\begin{figure} 
\typeout{*** EPS figure 2bc}
\begin{center}
\includegraphics[scale=0.3,angle=0]{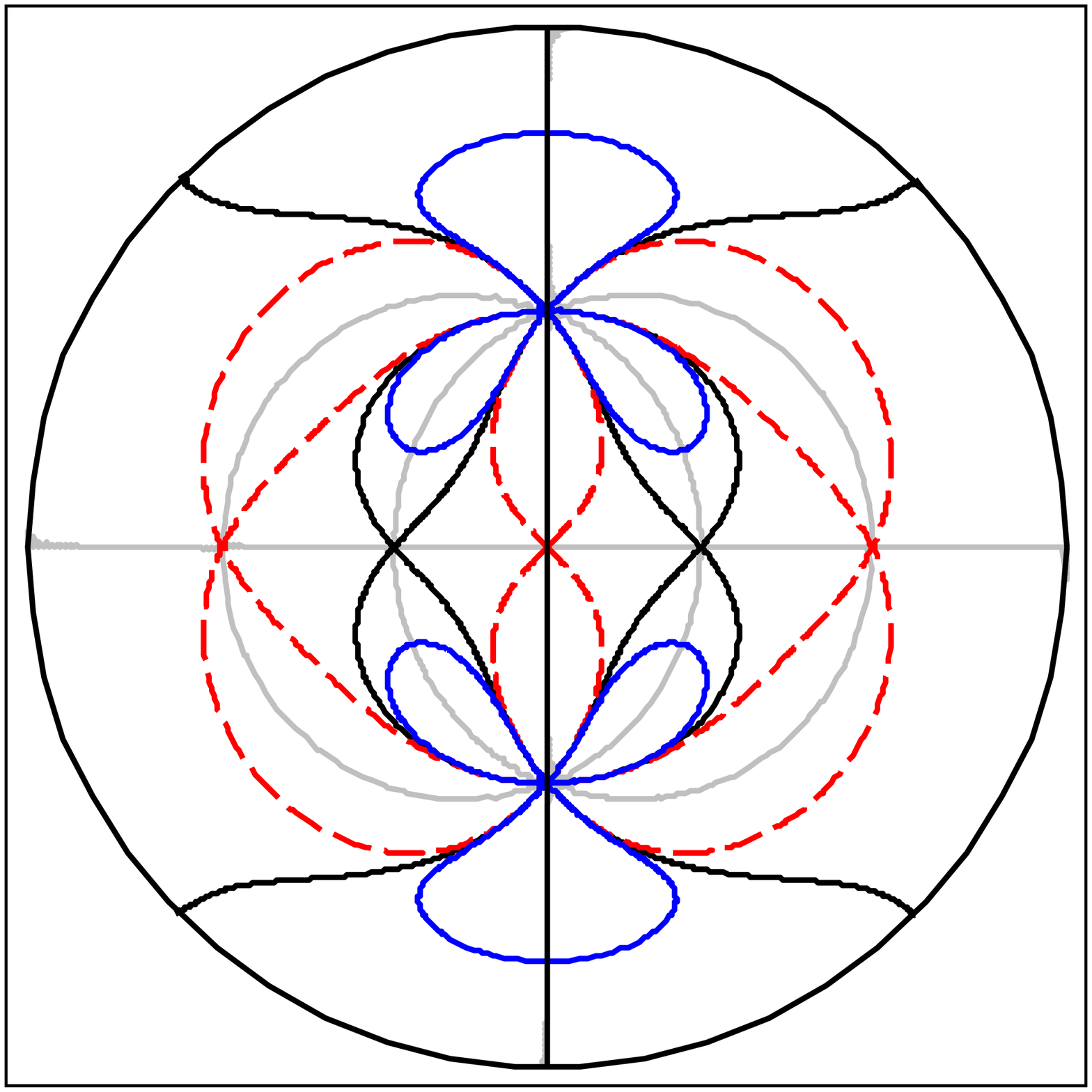} \qquad 
\includegraphics[scale=0.3,angle=0]{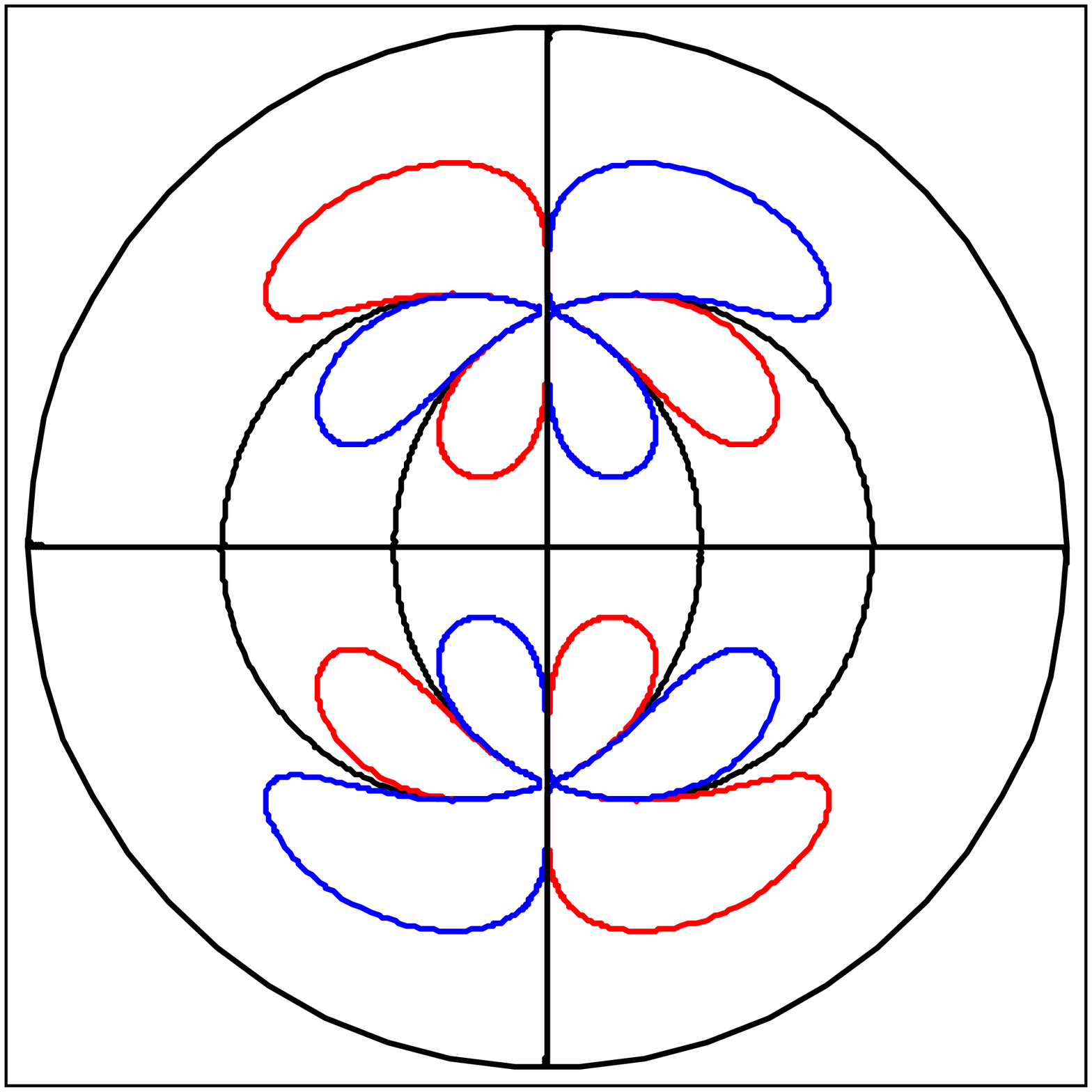}
\end{center}
\caption{Contours for the values $-1,0,1$ of the real (left) and imaginary (right) parts of the speciality index function $\mathcal{S}(\mu)$ are plotted using compactified polar coordinates in the $\mu$ complex plane centered on the point $(-1/2,0)$. The imaginary part shows clearly the two circles $|\mu+1/2|=1=|\mu| $ and the vertical line through their intersection points,  and which are also shown in gray in the real part plot, which divide the complex plane up into six equivalent regions under the symmetric group. The imaginary part is 0 on these curves and along the real axis.
$\mathcal{S}(\mu)$ has additional reflection symmetries about the center. The dashed curve in the real part plot is where the real part is 1, and its intersection with the gray curves  $\Im(\mathcal{S})=0$ give the speciality points where $\mathcal{S}=1$, namely, three points along the real axis.
}
\label{fig:2bc}
\end{figure}

To get a better feeling for the symmetries of these graphs, one can introduce polar coordinates centered on the point $(-1/2,0)$ and bring infinity in to the unit circle centered at that point by compactifying the radial coordinate
$$
  u+1/2 = \tan(r\pi/2) \exp(i\theta) \,.
$$
Now implicitly plotting the contours where the real and imaginary parts of $\mathcal{S}(\mu)$ equal $-1,0,1$ yields Fig.~\ref{fig:2bc}.


The original eigenvalue permutation symmetries of this function correspond to the following changes in $u$ associated with the three basic transpositions
\begin{eqnarray}
&&T_{12}(\lambda_1,\lambda_2,\lambda_3)=(\lambda_2,\lambda_1,\lambda_3):\quad  
\mu\rightarrow \mu^{-1}\,,\nonumber\\
&&T_{23}(\lambda_1,\lambda_2,\lambda_3)=(\lambda_1,\lambda_3,\lambda_2):\quad 
\mu\rightarrow -(1+\mu^{-1})^{-1}\,,\nonumber\\
&&T_{13}(\lambda_1,\lambda_2,\lambda_3)=(\lambda_3,\lambda_2,\lambda_1):\quad 
\mu\rightarrow -(1+\mu)\,,
\end{eqnarray}
so $\mathcal{S}(\mu)$ is unchanged under each of these transformations. The fixed points of these three transformations are respectively $T_{12}$: $\mu=1,-1$, $T_{23}$: $\mu=0,-2$ and $T_{13}$: $\mu=-1/2,\mu^{-1}=0$, while the two circles $|\mu+1|=1=|\mu|$ and the line $\mu+1=0$ connecting their intersection points (the poles) are invariant submanifolds under their action. 
The two cyclic permutations are compositions of the transpositions, explicitly
\begin{eqnarray}
&&P_{231}(\lambda_1,\lambda_2,\lambda_3)=(\lambda_2,\lambda_3,\lambda_1): 
\mu\rightarrow -(1+\mu)^{-1}\,,\nonumber\\
&&P_{312}(\lambda_1,\lambda_2,\lambda_3)=(\lambda_3,\lambda_1,\lambda_2): 
\mu\rightarrow -(1+\mu^{-1})\,,
\end{eqnarray}
which both have the same fixed points $\mu=-\frac12\pm i\frac{\sqrt 3}{2}$, which are the poles of $\mathcal{S}(\mu)$. It should be noted that each of these transformations of the complex plane are combinations of inversions, reflections and translations.

The special types D and II have two eigenvalues which are the same, corresponding to one of the three eigenvalue ratios being 1, related to each other by the permutation symmetries
\begin{eqnarray}
 \lambda_2/\lambda_1=1:&&\ \mu=1\,,\phantom{-/2} \ (\lambda_1,\lambda_2,\lambda_3)\propto (1,1,-2)\,,
\nonumber\\
 \lambda_3/\lambda_1=1:&&\ \mu=-2\,,\phantom{/2}\ (\lambda_1,\lambda_2,\lambda_3)\propto (-2,1,1)\,,
\nonumber\\
 \lambda_3/\lambda_2=1:&&\ \mu=-1/2\,,\ (\lambda_1,\lambda_2,\lambda_3)\propto (1,-2,1)\,,
\end{eqnarray}
so $1=\mathcal{S}(1)= \mathcal{S}(-2)=\mathcal{S}(-1/2)$.
The zeros of $\mathcal{S}$ are obviously $\mu=0,-1$ and in the limit $\mu^{-1}\to0$ corresponding to a third zero at infinity, again all related to each other by the permutation symmetries. These correspond to another privileged triplet of values which occur when two of the eigenvalues are of opposite sign
\begin{eqnarray}
 \lambda_2/\lambda_1=-1:&&\ \mu=-1\,,\phantom{{}^{-1}}  (\lambda_1,\lambda_2,\lambda_3)\propto (1,-1,0)\,,
\nonumber\\
 \lambda_3/\lambda_1=-1:&&\ \mu=0\,,\phantom{-{}^{-1}} (\lambda_1,\lambda_2,\lambda_3)\propto (0,1,-1)\,,
\nonumber\\
 \lambda_3/\lambda_2=-1:&&\ \mu^{-1}=0\,,\phantom{-} (\lambda_1,\lambda_2,\lambda_3)\propto (1,0,-1)\,.
\end{eqnarray}

\section{The Kasner family}

The  Kasner metric is
\begin{equation}
\rmd s^2=-\rmd t^2+t^{2p_1}\rmd x^2+t^{2p_2}\rmd y^2+t^{2p_3}\rmd z^2\,,  \label{LLLL}
\end{equation}
where the so-called Kasner indices (constants) satisfy
\begin{equation}
p_1+p_2+p_3 = 1 = p_1^2+p_2^2+p_3^2 \label{constr}
\end{equation}
so that they form a one-parameter family of triplets which can assume values in the interval $[-\frac13,1]$.
The space of distinct 3-geometries corresponds to the quotient of this family of spacetimes by the discrete group of permutations of the spatial coordinates and can be represented by any one of the six intervals of ordered values of the triplets: $p_a\le p_b\le p_c$, $(a,b,c)=\sigma(1,2,3)$, $\sigma\in S_3$.
The Kasner family admits two distinct special subcases for which two of the $p_a$ indices are equal: 1) $p_a=p_b=0$, $p_c=1$  for which the spacetime is flat and therefore of Petrov type O  and  2) $p_a=-1/3$, $p_b=p_c=2/3$  corresponding to the Kasner locally rotationally symmetric  type $D$ solution, with $\mathcal{S}=1$.

The matrix of mixed spatial extrinsic curvature tensor  components
$(K^a{}_b)=-1/2 (g^{ac}\dot g_{cb})=-\delta^b{}_a\ln(g_{bb}^{1/2})\,\dot{} $ is diagonal
\begin{eqnarray}
  (K^a{}_b) &=& -t^{-1}\, {\rm diag}(p_1,p_2,p_3)\,,\\
    K^c{}_c &=& -t^{-1}(p_1+p_2+p_3)=-t^{-1} \,,\\
  (K^a{}_b)/K^c{}_c &=& {\rm diag}(p_1,p_2,p_3)\,,
\end{eqnarray}
from which it is clear that the Kasner indices represent the scale free part of this tensor's eigenvalues.
The Kasner index constraints come from the scale invariant Hamiltonian constraint in vacuum with zero spatial curvature
\beq\fl\qquad
  (K^a{}_b K^b{}_c - K^a{}_a K^b{}_b)/(K^c{}_c)^2  = \frac{p_1^2+p_2^2+p_3^2 -(p_1+p_2+p_3)^2}{(p_1+p_2+p_3)^2}=0 \ ,
\eeq
which forces the corresponding paths in the minisuperspace of inner products to be null geodesics of the deWitt metric, and from the trace of the spatial Einstein equations, which is an evolution equation for the square root of the spatial metric determinant $g^{1/2}=t^{p_1+p_2+p_3}=t$
\beq\fl\qquad
 0 =  g^{-1/2} (\dot K{}^a{}_a)\,\dot{} 
   =  g^{-1/2} (g^{1/2}) \,\dot{}\,\,\dot{}
   = (p_1+p_2+p_3)(p_1+p_2+p_3-1)t^{-2}\,.
\eeq
Of course these same two independent Einstein equations are also satisfied by $p_a=0$, which is just flat spacetime, not so interesting.

Expanding the square and cube of the constraint $p_1+p_2+p_3=1$ and simplifying using both constraints leads respectively to two other identities
\begin{eqnarray}
  && p_1 p_2 + p_2 p_3 + p_1 p_3 = 0\,,\qquad (\rightarrow  \frac1{p_1}+\frac1{p_2}+\frac1{p_3} =0)
\nonumber\\
  && p_1^3+p_2^3+p_3^3=1 + 3p_1p_2p_3\,.
\end{eqnarray}
The first (which is the scale invariant combination of the two Kasner constraints) shows that the three products of the Kasner indices (equivalently their reciprocals) parametrize the three diagonal entries of a diagonal tracefree matrix modulo the overall scale. Similar using the first of these two identities to replace $p_1p_2$ in their product and then using the linear constraint to eliminate $p_1+p_2$ leads to
\beq \label{eq:cubic2}
  p_1 p_2 p_3 = -p_3^2 (1-p_3)\qquad \mbox{or}\qquad  p_1 p_2 = -p_3 (1-p_3)\,.
\eeq

There are a number of parametrizations of this family of Kasner indices by a single independent parameter. For example, the obvious procedure is to solve the basic pair of contraints in terms of any one of the indices, say $p_1$. Eliminating $p_3$ using the linear constraint leads to a quadratic equation in $p_2$ with coefficients depending on $p_1$, whose obvious solution (backsubstituting to express $p_3$ in terms of $p_1$ as well)\cite{cbbp2,bbb} is
\beq\fl
 p_2= \frac12([1-p_1] \pm [(1-p_1)(1+3p_1)]^{1/2})\,, \quad
 p_3= \frac12([1-p_1] \mp [(1-p_1)(1+3p_1)]^{1/2})\,.
\eeq

On the other hand the spatially flat Kasner spacetime plays an important role as a limiting case in the dynamics of spatially homogeneous spacetimes with more complicated symmetry, which gives an extra context for choosing the parametrization conveniently. The
very different trigonometric parametrization quoted by Stephani et al \cite{ES} turns out to be related to Misner's approach to this dynamics, while the one introduced by Lifshitz and Khalatnikov describes extremely efficiently a particular situation in this dynamical arena.


\begin{figure}[ht] 
\typeout{*** EPS figure 00}
\begin{center}
\includegraphics[scale=0.8]{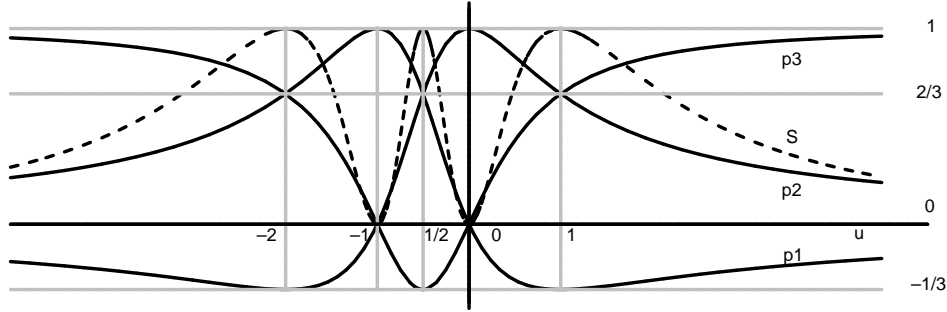}
\end{center}
\caption{The Kasner indices $(p_1,p_2,p_3)$ are plotted versus $u$, together with the speciality index $S$. The obvious reflection symmetry across $u=-1/2$ shows 3 pairs of corresponding intervals in this parametrization, but hides the equivalence of all 6 intervals corresponding to the permutations of $p_a\le p_b\le p_c$. $p_1$ and $p_2$ and the speciality index $S$ all have the limiting value 0 at $\pm\infty$ while $p_3$ goes to 1.
} 
\label{fig:3}
\end{figure}

The Lifshitz-Khalatnikov parametrization~\cite{LK,BK0,BK,BK2} represents the Kasner indices in terms of a single independent (real) parameter $u=u_{32}=p_3/p_2$ as follows
\beq\label{KL}
\fl\qquad
p_1=\frac{-u}{1+u+u^2}\,,\quad 
p_2=\frac{1+u}{1+u+u^2}\,, \quad 
p_3=\frac{u(1+u)}{1+u+u^2}\,.
\eeq
For a given ordered set of three distinct Kasner indices, there are six permutations that correspond merely to permuting the spatial coordinates, dividing the real $u$ axis up into six equivalent intervals (see Fig.~1) whose endpoints are the set
$\{-\infty,-2,-1,-1/2,0,1,\infty\}$.
Confining $u$ to the interval $u\in[1,\infty)$ leads to the
ordering of the Kasner exponents: $p_1\le p_2\le p_3$.
In general  $p_a\le p_b\le p_c$ for $u_{cb}\equiv p_c/p_b\in [1,\infty)$.

The Kasner index parametrization appears mysterious until one thinks about the two constraints on the indices. Both constraints fix the overall scale of the indices so that only their ratios can vary, so it is natural to pick one of those ratios as a new parameter, which can be done in six different ways. Picking $u= u_{32}=p_3/p_2$, for example, it is trivial to express all six possible ratios in terms of it. Eliminating $p_3=u p_2$ first, and then using the linear constraint to eliminate 
$p_1=1-p_2-u p_2=1-(1+u)p_2$, the quadratic constraint reduces to a quadratic equation in $u$
\beq
  [1-(1+u)p_2]^2 +p_2^2 + (u p_2)^2 =1
\eeq
with roots $p_2=0$ (obvious) and $p_2=u/(1+u+u^2)$, from which $p_1$ and $p_3$ are then obtained by backsubstitution. One can define six such parameters $u_{ab}=p_a/p_b$, for which $p_c\le p_b \le p_a$ when $(c,b,a)$ is a cyclic permutation of $(1,2,3)$, all of which are needed for application to the various pairs of axes in BLK-Mixmaster dynamics.
This is quite unremarkable, but on the other hand this simple class of parametrizations turns out to be quite remarkable in the BLK-Mixmaster dynamics, as will be discussed in detail below.

By straightforward evaluation  \cite{cbbp1,cbbp2}, one finds that the complex matrix $(Q^a{}_b)$ is already diagonal, real and purely electric (since the magnetic part of the Weyl tensor is zero), with the value 
\begin{eqnarray}
(Q^a{}_b) 
&=& -(E^a{}_b)
=-\dot K{}^a{}_b +K^a{}_c K^c{}_b \nonumber\\
&=& -t^{-2}{\rm diag}(p_1-p_1^2,p_2-p_2^2,p_3-p_3^2)
\nonumber\\
&=& t^{-2} {\rm diag}(p_2p_3,p_3p_1,p_1p_2) \,,
\end{eqnarray}
so Kasner is in general of Petrov type I and its eigenvalues are
\beq
(\lambda_1,\lambda_2,\lambda_3) = t^{-2}(p_2p_3,p_3p_1,p_1p_2)\,,
\eeq
so that
\beq\fl\qquad
\frac{\lambda_2}{\lambda_3} = \frac{p_3}{p_2} = u\,,\quad
\frac{\lambda_3}{\lambda_1} = \frac{p_3}{p_1} = -(1+u)\,,\quad
\frac{\lambda_1}{\lambda_2} = \frac{p_2}{p_1} = -(1+u^{-1})\,,
\eeq
hence $\mu_{23}=u$ is purely real for these spacetimes.
The permutations of the eigenvalues are locked to those of the Kasner indices, which were originally discussed by Misner \cite{misner69} using logarithmic diagonal metric variables adapted to the pure trace and tracefree decomposition of diagonal matrices 
\begin{eqnarray}\label{misner}\fl\quad
 (\ln(g_{11}^{1/2}),\ln(g_{22}^{1/2}),\ln(g_{33}^{1/2})
&=& (\beta^1,\beta^2,\beta^3) \nonumber\\\fl\quad
&=& \beta^0(1,1,1) + \beta^+ (1,1,-2) +\beta^- \sqrt{3}(1,-1,0)\,.
\end{eqnarray}
The Kasner indices are then functions of the ratios of the corresponding cotangent space (momentum) variables $(p_0,p_+,p_-)$.



\begin{figure}[ht] 
\typeout{*** EPS figure 4}
\begin{center}
\includegraphics[scale=0.6]{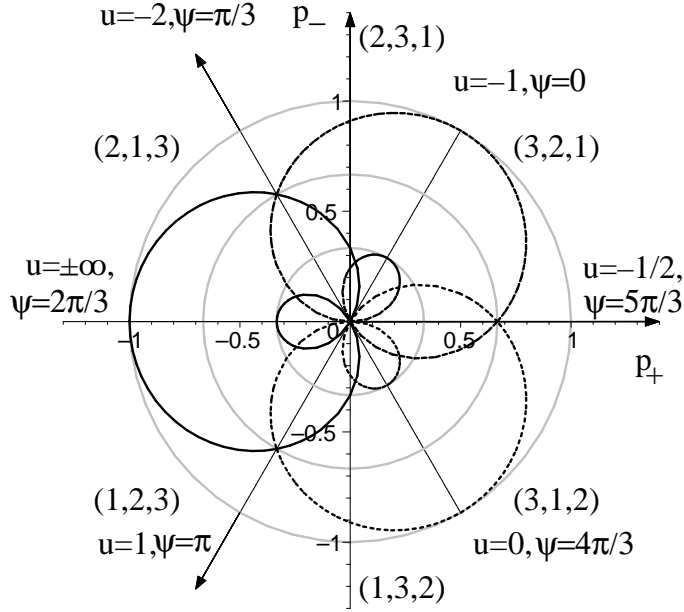}
\end{center}
\caption{The Kasner indices $(p_1,p_2,p_3)$, identified respectively by short dashed, long dashed and solid cardiods, are plotted 
in the ``Misner plane" of the $p_\pm$ momentum variables
along the radial direction versus the translated polar angle $\Psi$ measured counterclockwise from the half ray at an angle $\pi/3$ in the counterclockwise direction from the positive horizontal axis, where the negative Kasner indices lead to the internal loops. This establishes a correspondence of the Kasner indices and the parameters $u$ and $\Psi$ with the unit circle of directions in this $p_+$-$p_-$ plane, whose axes are adapted to the third direction $E_3$. The three circles have radii $1/3,2/3,1$, while each of the six sectors is labeled by a triplet $(a,b,c)$ indicating the ordering $p_a\le p_b\le p_c$ in that sector.
} 
\label{fig:4}
\end{figure}

To visualize the permutation symmetry in a symmetric way, it is useful to parametrize the indices using an angular polar coordinate in the diagonal tracefree matrix cotangent space following Stephani et al \cite{ES}
\begin{eqnarray}
\fl\qquad
  \left(p_1-\frac13,p_2-\frac13,p_3-\frac13\right)
  &=&\frac23\, \left(\cos\Psi,\cos(\Psi+\frac{2\pi}{3}),\cos(\Psi+\frac{4\pi}{3})\right)
\nonumber\\
\fl\qquad
  &=& -\frac13\,[ \cos\Psi\,(-2,1,1) + \sin\Psi\,\sqrt3 (0,1,-1) ]\,,
\end{eqnarray}
where here a permutation of Misner's tracefree diagonal matrix parametrization (\ref{misner}) occurs, here adapted to the first axis rather than the third axis as usual.
Fig.~\ref{fig:4} shows the Kasner indices plotted as the signed radial variable versus the polar angle $\Psi$ measured counterclockwise from the half ray along the direction rotated by $\pi/3$ counterclockwise from the positive horizontal axis. The triplets marking each of the six sectors give the ordering of the Kasner indices for that interval.


\begin{figure}
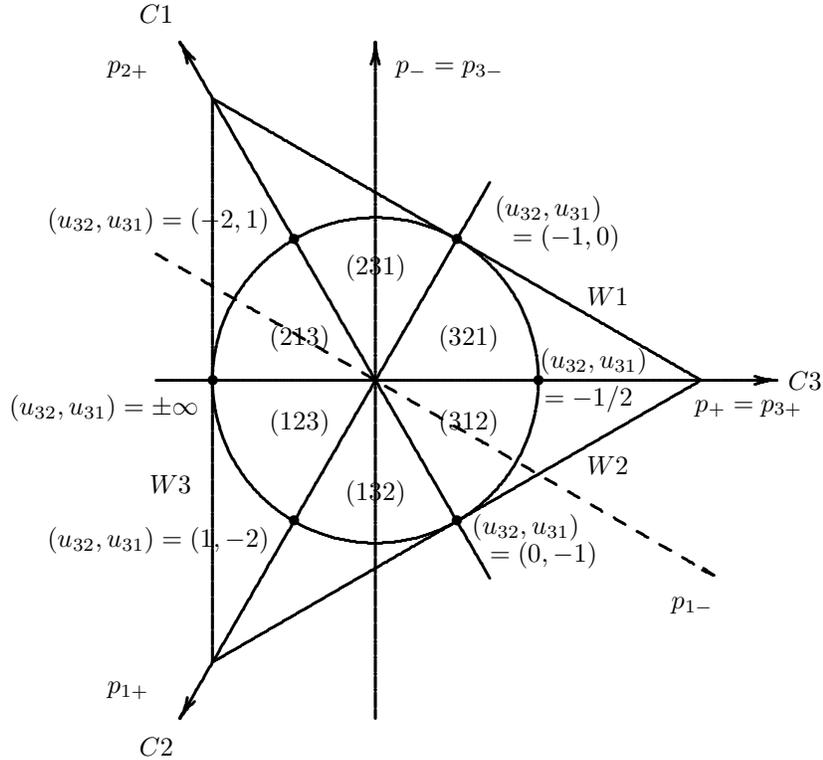

\typeout{figure pic}
$$ \vbox{
\beginpicture
\setcoordinatesystem units <0.75cm,0.75cm> point at  25 0  


\arrow <.3cm>  [.1,.4]    from  -1 0 to 10 0  
\arrow <.3cm>  [.1,.4]    from  2.887 -6 to 2.887 6  
 
\arrow <.3cm>  [.1,.4]    from  4.91  3.5 to -0.577 -6 
\arrow <.3cm>  [.1,.4]    from  4.91 -3.5 to -0.577 6 

\plot 0 -5 0 5 /        
\plot 0 -5 8.66 0 /     
\plot 0  5 8.66 0 /     

\put {\mathput{\bullet}}                at  0 0 
\put {\mathput{\bullet}}                at  5.774 0 
\put {\mathput{\bullet}}                at  4.33 -2.5 
\put {\mathput{\bullet}}                at  4.33  2.5 
\put {\mathput{\bullet}}                at  1.444   2.5 
\put {\mathput{\bullet}}                at  1.444  -2.5 

\circulararc 360 degrees from 0 0 center at 2.887 0 

\setdashes
\arrow <.3cm>  [.1,.4]    from  -1 2.244 to 8.887 -3.464   

\put {\mathput{p_+=p_{3+}}}    at  9.5 -0.5  
\put {\mathput{p_-=p_{3-}}}    at  4.2 5.5  
\put {\mathput{p_{2+}}}    at  -1.5  5.5
\put {\mathput{p_{1+}}}    at  -1.5 -5.5
\put {\mathput{p_{1-}}}    at  8.5 -4  

\put {\mathput{C3}}    at  10.5 0 
\put {\mathput{C1}}    at  -1  6.5 
\put {\mathput{C2}}    at  -1 -6.5 

\put {\mathput{W3}} [b]   at  -0.75 -2 
\put {\mathput{W1}}    at   7  1.5 
\put {\mathput{W2}}    at   7 -1.5 

\put {\mathput{(u_{32},u_{31})=\pm\infty}}  [tr]  at  -0.25 -0.25  
\put {\mathput{(u_{32},u_{31})}}  [lb]  at  5 2.8
\put {\mathput{\ \ =(-1,0)}}  [lb]  at  5 2.3
\put {\mathput{(u_{32},u_{31})}}  [l]  at  4.6 -2.6
\put {\mathput{\ \ =(0,-1)}}  [l]  at  4.6 -3.1
\put {\mathput{(u_{32},u_{31})}}  [lb] at  5.8 0.1
\put {\mathput{\, =-1/2}}  [lt]  at 5.8 -0.1
\put {\mathput{(u_{32},u_{31})=(-2,1)}}  [br]  at  1 2.6 
\put {\mathput{(u_{32},u_{31})=(1,-2)}}  [tr]  at  1 -2.6

\put {\mathput{(123)}}  [l]  at  1 -0.75
\put {\mathput{(213)}}  [l]  at  1  0.75
\put {\mathput{(321)}}  [l]  at  4  0.75
\put {\mathput{(312)}}  [l]  at  4 -0.75
\put {\mathput{(231)}}    at  2.887  2
\put {\mathput{(132)}}    at  2.887 -2

\endpicture}$$
\label{fig:5}
\caption{
The Kasner circle divided into six zones by its scattering properties with curvature walls $W1,W2,W3$ parallel to the sides of the equilateral triangle. The L-K parameters $u_{31}$ and $u_{32}$ describe the successive collisions of the system point with $W1,W2$ as it moves into and out of the corner $C3$. The momentum scattering can be imagined as taking place at the origin. For a case in which the system point is moving from wall $W2$ towards wall $W3$ while colliding with wall $W1$ as time increases, if the final state is in zone (123): $u\in (1,\infty)$, the initial state is in zone (231)$\cup$(213): $u\in(-\infty,-1)$, with a collision map which is simply $u\to-u$. In the increasing time direction, wall $W1$ is moving towards the corner $C1$ with half the speed of the system point, so even if the system point is moving at an angle up to 30 degrees from the direction of the wall $W1$ (dashed axis) in zone (213), a collision still occurs. The Kasner dynamics goes in the reverse time direction.
}
\end{figure}

\section{BKL-Mixmaster dynamics}

The Kasner indices themselves or the parameter $u$ or the angular parameter $\Psi$ may be thought of as parametrizing the unit circle in the $p_+$-$p_-$ plane of the two anisotropy momenta which coordinatize the cotangent plane to the tracefree metric matrices, themselves parametrized by the Misner \cite{misner69} variables $\beta^+$ and $\beta^-$, as explained in detail in \cite{unified}. Letting $(\hat p_+,\hat p_-)$ be the restriction of these variables to the unit circle, one has
\begin{eqnarray}
  (\hat p_+,\hat p_-) 
&=& (\cos(\Psi+\pi/3),\sin(\Psi+\pi/3))
\nonumber\\
&=& -(u^2+u+1)^{-1} ([u+1/2]^2-3/4,\sqrt{3}[u+1/2]) \,.
\end{eqnarray}
Since the DeWitt metric on the diagonal metric matrices is conformally Minkowskian (proportional to $-d\beta_0{}^2 + (d\beta^+)^2 + (d\beta^-)^2$, one can identify the unit circle in the anisotropy momentum space with the unit circle in the anisotropy velocity space, representing the space of directions in the $\beta^\pm$ anisotropy plane. This ``Kasner circle" represents the asymptotic Kasner states in a scattering off a curvature potential in spatially homogeneous dynamics idealized as a collision with a moving potential wall in this anisotropy plane.


\begin{figure}
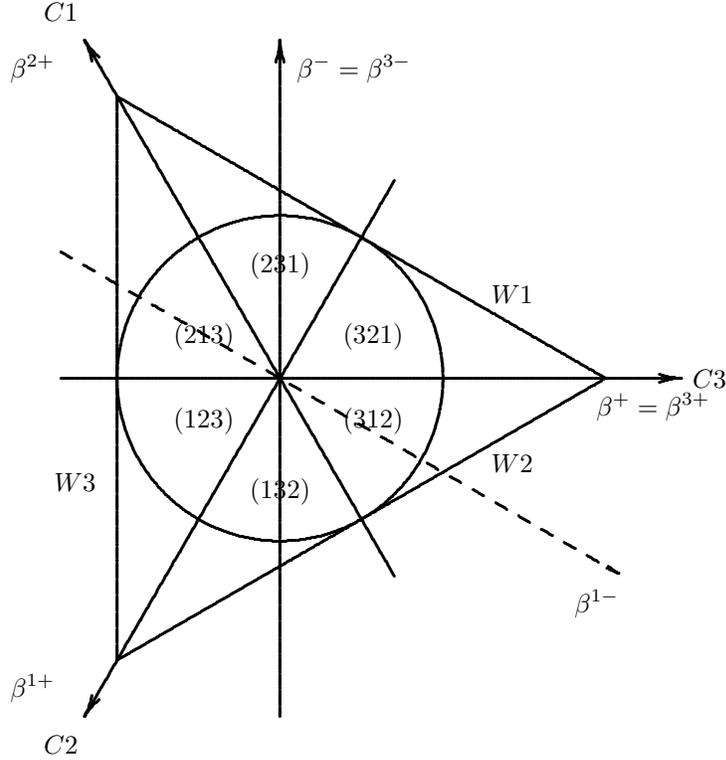

\typeout{figure pic2}
$$ \vbox{
\beginpicture
\setcoordinatesystem units <0.75cm,0.75cm> point at  25 0  


\arrow <.3cm>  [.1,.4]    from  -1 0 to 10 0  
\arrow <.3cm>  [.1,.4]    from  2.887 -6 to 2.887 6  
 
\arrow <.3cm>  [.1,.4]    from  4.91  3.5 to -0.577 -6 
\arrow <.3cm>  [.1,.4]    from  4.91 -3.5 to -0.577 6 

\plot 0 -5 0 5 /        
\plot 0 -5 8.66 0 /     
\plot 0  5 8.66 0 /     

\circulararc 360 degrees from 0 0 center at 2.887 0 
\setdashes
\arrow <.3cm>  [.1,.4]    from  -1 2.244 to 8.887 -3.464   

\put {\mathput{\beta^+=\beta^{3+}}}    at  9.5 -0.5  
\put {\mathput{\beta^-=\beta^{3-}}}    at  4.2 5.5  
\put {\mathput{\beta^{2+}}}    at  -1.5  5.5
\put {\mathput{\beta^{1+}}}    at  -1.5 -5.5
\put {\mathput{\beta^{1-}}}    at  8.5 -4  

\put {\mathput{C3}}    at  10.5 0 
\put {\mathput{C1}}    at  -1  6.5 
\put {\mathput{C2}}    at  -1 -6.5 

\put {\mathput{W3}} [b]   at  -0.75 -2 
\put {\mathput{W1}}    at   7  1.5 
\put {\mathput{W2}}    at   7 -1.5 

\put {\mathput{(123)}}  [l]  at  1 -0.75
\put {\mathput{(213)}}  [l]  at  1  0.75
\put {\mathput{(321)}}  [l]  at  4  0.75
\put {\mathput{(312)}}  [l]  at  4 -0.75
\put {\mathput{(231)}}    at  2.887  2
\put {\mathput{(132)}}    at  2.887 -2

\endpicture}$$
\label{fig:6}
\caption{
The anisotropy plane interior to the three curvature walls $W1,W2,W3$ forming an equilateral triangle is divided into six zones by the perpendicular bisectors of the sides, conveniently labeled by the corresponding six Kasner circle intervals. In turn these pair up into three corner zones $C1$, $C2$, $C3$.
The L-K parameters $u_{31}$ and $u_{32}$ of the previous figure describe the successive collisions of the system point with $W1,W2$ as it moves into and out of the corner $C3$.
}
\end{figure}

As explained in detail elsewhere \cite{unified}, a collision with each of the three potential walls arranged with equilateral triangle symmetry is idealized as an exact Bianchi type II solution for the diagonal metric components. In terms of the $u=u_{32}$ Lifshitz-Khalatnikov parametrization of the Kasner circle, with  the future time asymptotic value $u_\infty\in (1,\infty)$ and past time asymptotic value  $u_{-\infty}\in (-\infty,-1)$, a collision corresponds to the system point moving from wall 2 towards wall 3 while colliding with wall 1. The simple geometric reflection which takes place in the rest frame of the moving wall is boosted when described in the nonmoving frame.

The pictorial representation of the Kasner map in terms of a collision with a curvature potential wall in this anisotropy plane is complicated by the fact that the dynamics is reversed in time (as one approaches an initial cosmological singularity, the cosmic time $t$ decreases) while the potential wall is moving, so the analogy is more like pinball than billiards since the moving wall acts like a flipper in transferring momentum to the system point. 
Fig.~6 
indicates the orientations of the three equilateral triangle curvature walls of BLK-Mixmaster dynamics relative to the Kasner circle  (not the walls themselves, which move in the $\beta^\pm$ plane, see 
Fig.~7. 

Each of these curvature walls moves with a speed which is half the speed of the system point itself, allowing a relative motion of the system point of up to 30 degrees away from a direction parallel to a straight line wall and still collide with the wall as the walls move inward as the cosmic time increases. For example, in 
Fig.~6 
the incoming unit vector with values in regions (2,3,1) and (2,1,3) ranging from direct normal incidence ($u=-1$) to moving away at an angle 30 degrees from the normal ($u=-\infty$) is kicked by the wall into the region (1,2,3). 

Of course it is not enough to look at the diagram in velocity/momentum phase space. One must couple this with the actual position in the $\beta^\pm$ plane relative to the actual triangular walls. See 
Fig~7. 
The three bisector line segments of the equilateral triangle angles extending from the vertex to their common meeting point at the center divide up the triangle into three isoceles triangles, each of which has one of the three walls $W1$, $W2$, $W3$ as a base, while extending those bisectors through the center to the opposite side further divides these into six right triangles, one for each sector of the Kasner circle. These in turn can be matched up in pairs corresponding to the three corners $C1$, $C2$, $C3$. 
Fig.~6 
then describes the directions in the tangent space to a system point located in one of these six triangles.

Suppose the system point is in corner $C3$ with a final direction in the Kasner interval (1,2,3), namely $u_{32}=u\in (1,\infty)$, so that it is escaping towards a collision with wall $W3$ after a collision with wall $W1$. Its initial incoming direction was in the Kasner interval (2,1,3), coming from a collision with wall $W2$, whose final direction corresponds to $i_{32}=-u$, $u_{31}=u-1$. If $u-1>0$, then the initial state for this collision will have $u_{31}=-(u-1)$, $u_{32}=u-2$. Working backwards in time, we are describing a Kasner era consisting of a series of incoming-outcoming Kasner states (epochs) which continue until $u-n<0$, and the system point escapes corner $C3$ moving on to a Kasner era bouncing between the two walls of another corner. To graphically record this dynamics in the $\beta^\pm$-plane, one rescales the wall position variables in order to fix them in place, giving an equilateral triangle billiard representation of the more complicated moving wall bouncing of the system point, thus removing the scale in anisotropy space. See the figures in Chapter 13 of Ryan and Shepley \cite{ryanshepley}, taken from Ref.~\cite{mosermatznerryan}.

The transitions in the scalefree part of the Weyl tensor eigenvalues during each collision with a curvature wall are remarkably simple in terms of the L-K parameters. For example, if $u\ge2$ then in the direction towards the initial singularity a collision with wall $W1$ in corner $C3$ (the standard way in which the Ksner map is usually stated for ordered Kasner indices) leads to
\begin{eqnarray}
  (\lambda_1,\lambda_2,\lambda_3) 
= \lambda_2 (u, 1, -1-u) 
&\rightarrow& \lambda_2^\prime ( -u,1,-1+u)\nonumber\\
&=& \lambda_1^\prime (1,u-1,-u)\,.
\end{eqnarray}
The first eigenvalue ratio is just reversed in sign, and the final ratio compensates to keep the sum zero, with respect to the same eigenvector basis. Transposing the basis as well leads to the unit subtraction of the old ratio to form the new ratio appropriate for the new basis.

\section{Concluding remarks}

While not a very fundmental revelation, finally the mysterious Lifshitz-Khalatnikov Kasner index parametrization is understood in a larger context than the narrow one in which it was introduced without explanation in order to do its job. It is satisfying to pull on a bit of mathematics and discover that it is hiding a lot more than meets the eye initially. Many of us were attracted to relativity in part for its mathematical beauty, and these little gems are always a joy to discover.



\end{document}